\documentclass[fleqn,twoside]{article}
\usepackage{espcrc2}
\usepackage{epsf}

\def\prd#1{Phys.\ Rev.\ {\bf D#1}}

\title{Heavy-light meson decay constants with $N_f=3$}
\author{
MILC Collaboration:
C.~Bernard,\hskip-0.03in
\address{{\vskip-0.10in{\hskip 0.07in Department of Physics, Washington
University, St.~Louis, MO 63130, USA}}} 
T.~Burch,\hskip-0.03in
\address{Department of Physics, University of Arizona, Tucson, AZ 85721, USA} 
S.\ Datta,\hskip-0.03in
\address{Department of Physics, Universit\"at Bielefeld, Bielefeld, Germany} 
C.~DeTar,\hskip-0.03in
\address{Physics Department, University of Utah, Salt Lake City, UT 84112, USA}
Steven~Gottlieb,\hskip-0.03in
\address{Department of Physics, Indiana University, Bloomington, IN 47405, USA} 
\hskip-0.03in\thanks{presented by S.\ Gottlieb}
E. Gregory,${^b}$
\advance\baselineskip -2pt
Urs~M.~Heller,\hskip-0.03in
\address{CSIT, Florida State University, Tallahassee, FL 32306-4130, USA} 
R.~Sugar\hskip0.005in
\address{Department of Physics, University of California, Santa Barbara, CA
93106, USA} 
and
D.~Toussaint$\null^{\rm b}$ 
\advance\baselineskip -2pt
       }

\begin{document}

\begin{abstract}
During the past year the MILC Collaboration has continued its study of
heavy-light meson decay constants with three dynamical quarks.  
Calculations have been extended to a second lattice spacing of about
0.09 fm.  At this lattice spacing, there are results in the quenched
approximation and for three sets of dynamical quark mass: $m_l=m_s$;
$m_l=0.4 m_s$ and $m_l=0.2 m_s$, where $m_l$ is the light mass for the
$u$ and $d$ quarks and $m_s$ is the strange quark mass.  At the coarser
lattice spacing, for which results were presented at Lattice 2001, statistics
have been increased for two sets of quark masses and three additional
sets of quark masses have been studied, giving a total of eight combinations
used to interpolate between the quenched and chiral limits.
When these calculations are completed, we can study the decay constants taking
into account both chiral and continuum extrapolations.

\end{abstract}

\maketitle

\section{INTRODUCTION}

We are extending a calculation of heavy-light meson decay constants
with three flavors of dynamical quarks that was begun last year \cite{LAT01}.
In addition to increasing statistics on some runs, we have additional
mass combinations for $a=0.13$ fm and new dynamical quark runs with
$a=0.09$ fm.  Runs for the coarser lattice spacing are completed, but
running will continue for the finer lattice spacing, so that we can
begin to understand the continuum limit.

Dynamical gauge configurations are generated using the Asqtad action 
\cite{ASQTAD}.
For the heavy-light mesons, we use tadpole-improved
clover valence quarks and operators that are improved according to the
Fermilab formalism \cite{EKM}.  
For each ensemble of dynamical quark configurations,
we use five light and five heavy valence quark masses.  The masses and
decay constants are interpolated or extrapolated as explained below to 
get physically relevant values with the overall scale set by the $\rho$ mass.

\section{PROGRESS SINCE LATTICE 01}
Last year we had completed running for $a=0.13$ fm on quenched configurations,
on a two flavor ensemble, and for four combinations of light and strange
quark dynamical masses.  A fraction of another set was run.
We have now completed eight sets of dynamical masses with $a=0.13$ fm.
Our quenched run for $a=0.09$ fm included about 15\% 
of the configurations there,
and we were generating configurations with two sets of dynamical masses.
We soon completed analysis of every-other quenched lattice and have made
substantial progress on three dynamical ensembles.  The table below contains
a summary of our running.  

\begin{table}[th]
\begin{tabular}{cccc}
\noalign{\hrule}
\small
  dynamical & $\beta$ & configs.    & configs.  \\
\noalign{\vskip -.07truein}
              $am_{u,d}/am_s$ &      & generated   & analyzed  \\
\noalign{\hrule}
\noalign{\hrule}
\noalign{\smallskip}
\noalign{\hrule}
\noalign{\hrule}
\noalign{\smallskip}
\multicolumn{4}{c}{$a=0.13$ fm; $20^3\times64$}\\
\noalign{\hrule}
\noalign{\smallskip}
    $\infty$/$\infty$ & 8.00 & 408 &  290  \\
\noalign{\hrule}
\noalign{\smallskip}
    0.02/$\infty$ & 7.20 & 411 &  411  \\
\noalign{\hrule}
\noalign{\smallskip}
    0.40/0.40   & 7.35 & 332 &  324  \\
    0.20/0.20   & 7.15 & 341 &  341  \\
    0.10/0.10   & 6.96 & 340 &  340 \\
    0.05/0.05   & 6.85 & 425 &  425  \\
    0.04/0.05   & 6.83 & 351 &  347  \\
    0.03/0.05   & 6.81 & 564 &  563  \\
    0.02/0.05   & 6.79 & 486 &  486  \\
    0.01/0.05   & 6.76 & 407 &  399  \\
\noalign{\hrule}
\noalign{\hrule}
\noalign{\smallskip}
\noalign{\hrule}
\noalign{\hrule}
\noalign{\smallskip}
\multicolumn{4}{c}{$a=0.09$ fm; $28^3\times96$}\\
\noalign{\hrule}
\noalign{\smallskip}
    $\infty$/$\infty$ & 8.40 & 417 &  200  \\
\noalign{\hrule}
\noalign{\smallskip}
    0.031/0.031   & 7.18 & 336 &  163  \\
    0.0124/0.031   & 7.11 & 370 &  120  \\
    0.0062/0.031   & 7.09 & 176  &  48  \\
\noalign{\hrule}

\end{tabular}
\end{table}

\section{ANALYSIS OF RESULTS}


The analysis of the heavy-light decay constants involves a number of steps:

1. {fit light pseudoscalar masses}

2. {perform chiral fit of pseudoscalar masses to determine $\kappa_c$ 
}


3. {fit light vector meson masses}

4. {perform chiral fit of vector meson masses and determine $m_{u,d}$
and $a$ to get physical $m_\pi/m_\rho$ and $m_\rho$}

5. {determine $m_s$ from mass of $\bar s s$ pseudoscalar state assuming
linear chiral mass relation}

6. {fit heavy-light channels to determine masses and decay amplitudes
}


7. {extrapolate or interpolate results in light quark mass to $m_{u,d}$
or $m_s$, respectively
(see Fig.~1).  
One can see that the confidence level of the nonlinear fit
is much better than that of the linear fit.  The curvature of the former has
the same sign as that expected from chiral logarithms as 
recently suggested \cite{CHIRALLOGS};  
however, our data has considerably less curvature than the lowest order
chiral logarithms would predict, requiring
higher-order terms and, perhaps, a modification of the logarithms due to
`taste' breaking or partial quenching.}

\begin{figure}[t]
\epsfxsize=0.99 \hsize
\epsffile{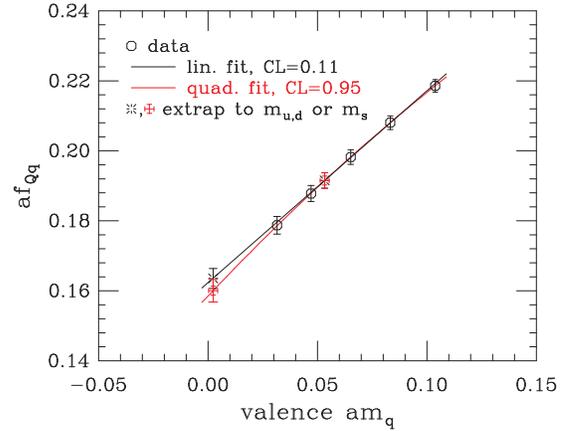}
\vspace{-38pt}
\caption{Chiral extrapolation of $f_B$
showing both linear and non-linear fits for $\beta=6.79$, $\kappa_Q=0.08$.}
\end{figure}

8. {after removal of perturbative logarithms, fit $f_{Qq}\sqrt{m_{Qq}}$ 
to a power series in $1/m_{Qq}$ and interpolate to $B$, $B_s$, $D$ and $D_s$ 
meson masses
}


9. {put the perturbative logarithm back and use 
the heavy-light axial-vector current
renormalization constant to get the renormalized decay-constant}

Unfortunately, the axial-vector renormalization constant has not yet been
calculated either perturbatively or nonperturbatively.  We assume that in the
quenched approximation our current results at $a=0.13$ fm
with an improved action agree with
the continuum limit of our previous calculation using Wilson and clover
quarks and the Wilson gauge action.  This was explained in more detail in
Ref.~\cite{LAT01}.

After the above procedure is done on each ensemble, we have a partially
quenched result at a particular value of dynamical $m_\pi/m_\rho$.
We then plot these results as a function of $(m_\pi/m_\rho)^2$ to perform
a chiral extrapolation.  This is demonstrated for $f_B$ in Fig.~2.

\begin{figure}[t]
\epsfxsize=0.99 \hsize
\epsffile{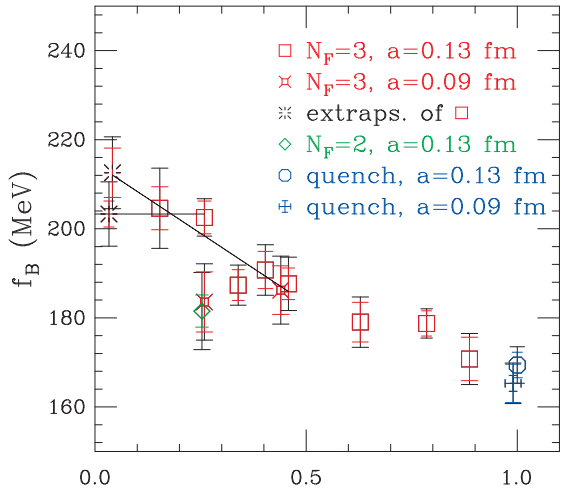}
\vspace{-38pt}
\caption{$f_B$ as a function of $(m_\pi/m_\rho)^2$.}
\end{figure}

It is worthwhile to plot the ratio of $B_s$ and $B$ meson decay constants
since many of the systematic errors are common, and a good deal of the
uncertainty from the renormalization constants drops out.  
Figure 3 shows
the ratio along with a constant chiral extrapolation in the dynamical mass.


\begin{figure}[t]
\epsfxsize=0.99 \hsize
\epsffile{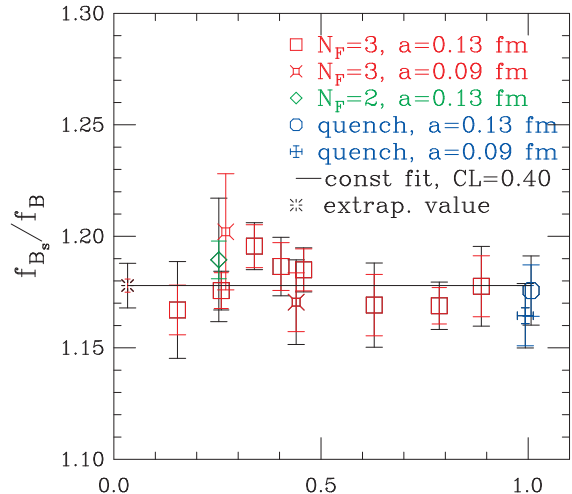}
\vspace{-38pt}
\caption{$f_{B_s}/f_B$ as a function
of $(m_\pi/m_\rho)^2$.}
\end{figure}



\section{CONCLUSIONS}

Our preliminary three flavor results and error estimates are slightly changed from
last year.  With higher statistics the quenched results at the finer lattice
spacing are in good agreement with the coarser spacing for the $B$ mesons.  
We find:

$$ f_B/f_B^{\rm quench} [m_\rho {\rm scale} ] = 1.23(4)(6) , $$

$$ f_{B_s}/f_B= 1.18(1)(^{+4}_{-1})  .$$

If we take the MILC continuum quenched value ($m_\rho$ scale) for
$f_B$ (169 MeV) as given, our value $f_B/f_B^{\rm quench}=1.23$
gives $f_B\approx207$ MeV with 2+1 flavors.  An independent determination
of $f_B$ awaits the calculation of the axial current renormalization factor.
The errors are detailed in the table below.

Better understanding of the chiral logs is necessary
and may change the $f_B$ chiral extrapolation beyond the error estimated here.

\begin{table}[ht]
\begin{tabular}{cccc}
   &$\ \ f_B/f_B^{\rm quench}$ & $f_B$ & $f_{B_s}/f_B$\\
\noalign{\hrule}
\noalign{\hrule}
{\bf prelim. result}&1.23& $-$ & 1.18 \\
\noalign{\hrule}
{\bf stat. error}& 3\% & 3\% & 1\% \\
\noalign{\hrule}
{\bf val. $\chi$ extr. err.}& 3\% & 3\% & ${}^{+3}_{-0}$\% \\
\noalign{\hrule}
{\bf dyn. $\chi$ extr. err.}& 2\% & 2\% & $<1$\% \\
\noalign{\hrule}
{\bf perturb. err.}& 2\% & ?? & $\ll$1\% \\
\noalign{\hrule}
{\bf discret. err.}& $<$3\% &3\%? & $<$1\% \\
\noalign{\hrule}
\noalign{\hrule}
\vspace{-10pt}
\end{tabular}
\end{table}

\end{document}